\documentclass[conference]{IEEEtran}
\IEEEoverridecommandlockouts
\usepackage{cite}
\usepackage{amsmath,amssymb,amsfonts}
\usepackage{algorithmic}
\usepackage{graphicx}
\usepackage{textcomp}
\usepackage{xcolor}
\def\BibTeX{{\rm B\kern-.05em{\sc i\kern-.025em b}\kern-.08em
    T\kern-.1667em\lower.7ex\hbox{E}\kern-.125emX}}
\begin{document}

\title{An Information-Theoretic Framework for Identifying Age-Related Genes Using Human Dermal Fibroblast Transcriptome Data\\

}
\author{\IEEEauthorblockN{ Salman Mohamadi}
\IEEEauthorblockA{\textit{Computer Science and Electrical Engineering} \\
\textit{West Virginia University}\\
Morgantown, WV, USA \\
Sm0224@mix.wvu.edu}
\and
\IEEEauthorblockN{Donald A. Adjeroh}
\IEEEauthorblockA{\textit{Computer Science and Electrical Engineering} \\
\textit{West Virginia University}\\
Morgantown, WV, USA \\
donald.adjeroh@mail.wvu.edu}
}

\maketitle

\begin{abstract}

    Investigation of age-related genes is of great importance for multiple purposes, for instance,  improving our understanding of the mechanism of ageing, increasing life expectancy, age prediction, and other healthcare applications. In this work, starting with  a set of 27,142 genes, we develop an information-theoretic framework for identifying genes that are associated with aging by applying  unsupervised and semi-supervised learning techniques on human dermal fibroblast gene expression data. 
    First, we use unsupervised learning and apply  information-theoretic measures to identify key features for effective representation of gene expression values in the transcriptome data. Using the identified features, we perform clustering on the data. Finally, we apply semi-supervised learning on the clusters using different distance measures to identify novel genes that are potentially associated with aging. Performance assessment for both unsupervised and semi-supervised methods show the effectiveness of the framework. 
    
\end{abstract}
\vspace{-.2cm}

\section{Introduction}
\vspace{-.2cm}
Aging is a complex process, and is often associated with the process of increasing physiological decline, increased risk of illness, and increased mortality. This process could manifest in terms of various types of decline in physiological function, such as transcriptional decline \cite{b1}. Early studies on ageing confirm the contribution of genes to ageing, for instance, in some species such as mouse or dog even a few number of gene mutations resulting from natural selection can lead to a significant extension of life span \cite{b2}. From the perspective of cell biology, ageing can be interpreted in terms of variations in gene expression due to both the environment and the genetic makeup. While some studies have showed that the age-associated genes are different among both different tissues and organisms, further analysis on large datasets identified some general common patterns\cite{b1}. Accordingly, certain changes in the level of expression, or in the patterns of expression of a subset of genes could imply that those  genes are associated with the process of ageing \cite{b3}. Furthermore, for a given gene, certain patterns of changes in its expression levels across the human life span could be an indicator of causal or associative relationship between the gene and the ageing process. In parallel with advanced experimental investigations in biological wet-laboratories, detection of the underlying patterns of expression calls for development of novel computational techniques  for the huge gene expression datasets.

The process of ageing of an organism (also called senescence \cite{b4}), includes nine hallmarks including stem cell reserves depletion, epigenetic alterations, genomic instability, mitochondrial dysfunction and a few other halmarks \cite{b5}. These hallmarks affect and are influenced by transcriptional changes \cite{b5}. 
Studies on the mechanism of these changes could provide some answers on whether these changes  are beneficial or detrimental. Accordingly, some studies show that ageing process can be regulated by therapeutic steps for longevity to amplify the protective changes and eliminate the detrimental changes \cite{b6,b1}.   

With an emphasis on the gene expression changes as one of the main elements of ageing process, there are different sets of theories including evolutionary theories and passive theories, trying to explain the hidden mechanisms responsible for gene expression changes associated with senescence \cite{b7}. For instance, a study on gene expression data of mice confirmed the ageing theory based on cellular metabolic stability \cite{b8}. Although an organism's case-by-case differences in the speed of ageing can be caused by efficiency in cellular and tissue metabolism, the general process of ageing for an organism is associated with changes in its gene expression. From a general perspective, recent studies contrast gene expression changes in terms of  global scale versus local scale. This  can be interpreted as redirecting efforts toward identifying certain subsets of genes responsible for the ageing process of different cell types (such as brain, pancreatic, kidney, liver, muscle, eye, and cell), tissues, and even species \cite{b1}. Recent studies such as Haustead et al \cite{b10} and Stegeman et al \cite{b1} suggest that less than 2${\%}$ of genes in human skin are identified with age-related changes in expression levels. See also  \cite{b10,b11}. 
Motivated by the foregoing, our goal in this work is to further investigate such subset of genes using advanced machine learning techniques.

In this paper, we develop an information-theoretic computational framework to analyze a dataset of  human dermal fibroblasts transcriptome data with gene expression values for 143 individuals. The framework is used to identify hidden patterns of gene expression associated with aging, for some subset of the 27,142 genes. Initially, an unsupervised approach is used to extract important features using information-theoretic measurements, based on which the genes are grouped into clusters with respect to ageing. The clusters are then further analyzed and refined using semi-supervised learning to identify potentially new age-related genes, exploiting  prior information from a small subset of genes known to be associated with ageing.



 The paper is organized as follows. The next section reviews related work on gene selection for the ageing process. 
 The next section presents our three-step methodology: (1) information theoretic feature measurements, (2) clustering, and (3) semi-supervised learning.  Section \ref{results} provides an evaluation of the proposed methods.  Section \ref{conclusion} provides a brief discussion and conclusion.

  \vspace{-.2cm}
\subsection{Prior Work}
 \vspace{-.2cm}
Machine learning algorithms have surpassed classical signal processing methods and have gained dramatic attention of computational biologists due to multiple advantages such as accuracy, robustness and good generalization \cite{b12,b27,b30,b35,b37}. Specifically, machine learning has showed its capability for genomic data processing. One good example is the study of mechanisms associated with ageing through gene expression profiling in order to estimate the age \cite{b13}.
The study of the mechanism of ageing, and in particular, the identification of age-related gene have attracted a lot of attention over the years \cite{b14}, especially given its implications on the possibility for developing therapeutic approaches to slow down or control the human ageing process  \cite{b15,b16}. It is hoped that such  manipulations can be implemented at  different scales, from whole body to organs to  small tissues  \cite{b1}. However, most attention has been directed toward costly experimental studies in laboratories. On the other hand, more recently, computational methods on selecting age-related genes have gained more attention, given their potential for facilitating more cost-effective understanding of mechanism of ageing, for instance, by providing a guide to experimental researchers for more focused investigations. For instance, 
Srivastava et al \cite{b17} employed machine learning tools to predict genes associated with age-related diseases. 
De Maclhaes et al \cite{b18}  investigated age-related genes and provided a meta-analysis of work on gene expression data of mice, rats and humans. They considered the noisy nature of gene expression data, and studied how to associate transcriptional changes with the ageing. Their results identified a set of 56 genes as overexpressed over the life span, and another set of 17 as underexpressed. Uddin et. al. \cite{b19} performed an analysis age-related variations in the gene expression data of porcine tissues in order to 
identify some reference genes. They identified 9  genes as the reference genes, and used these for data normalization of the gene  expression data.

Avelar et al \cite{b20} perform integrative computational analysis to select and validate a subset of age-related genes. Based on their comparative study they claim that human tissues' gene over-expression  with age can be used to characterize ageing and identifying age-related genes as well as age inhibiting  genes.

\begin{figure}[t]
\begin{center}
\includegraphics[width=1.\linewidth]{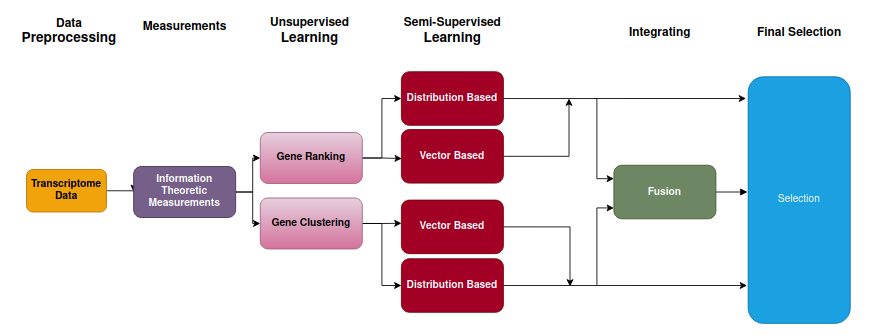} \rule{0.1\linewidth}{0pt}
   
\end{center}
\vspace{-1.2cm}
   \caption{Proposed pipeline for identifying age-related genes}
\label{fig:001_PL}
\label{fig:PL1}
\end{figure}

   

 \vspace{-.1cm}
\section{Methodology}
 \vspace{-.2cm}
 In this section, we discuss our dataset from \cite{b13} and introduce our computational framework. Briefly, besides a list of genes identified to be age-related by previous studies \cite{b14}, we have a dataset of human dermal fibroblast gene expression values for 27,142 genes across 143 individuals with age ranging from less that 1 year to 94 years old \cite{b13}. We  perform  a cascade of coarse and fine-grained gene selection using both unsupervised and semi-supervised learning. In the unsupervised approach, we first represent the expression values for each gene using  handcrafted information-theoretic measurements. We then apply a clustering algorithm on the new representation to cluster the genes into two groups. The idea behind the binary clustering is to group the genes based on their potential association with ageing. One rationale here is that, as the vector of expression values for each gene are represented over age (time axis), binary clustering on these raw vectors could expose the discriminating dynamic (behavior) of expression values with respect to aging. Another rationale is that, rather than just using the raw data for the binary clustering, the information-theoretic measurements are performed with respect to aging, thus the binary grouping could help to provide the first level discrimination between genes that are associated with aging, and those that are not.
 
 We expect that age-related hidden pattern(s) of changes should  manifest in the form of variations of certain measurements such as entropy, Kullback-Leibler divergence and correlation between the expression values and time axis built upon the age range. The credibility of this idea will be assessed in by  contrasting experimental results using raw data with  those obtained using the information-theoretic measurements. Yet, because noise and  other possible unknown factors could be contributing to the variations in the indicated  measurements along with the age, we cannot be certain that these variations are only due to hidden patterns associated with the aging process. To address this issue, we refine the results from the unsupervised approach using semi-supervised learning. Thus, using a small set of previously identified genes known to be associated with ageing, we performing semi-supervised learning for final identification of age-related genes from the previously clustered genes.
 
  \vspace{-.2cm}
\subsection{Datasets}
 \vspace{-.2cm}
For this work, we use the human dermal fibroblast transcriptome data reported in  \cite{b13} which contains data on gene expression values. 
The  database of gene expression values consists of a matrix of gene expression values from 27,142 genes across 143 individuals. Each row contains the expression values from the genes across 143 subjects in order of age (from less than 1 to 94 years old). Each column captures the expression values for each of  27,142 different genes for a given individual. We note that, the expression values for any given gene in the data set are measured from different individuals (143 subjects with age ranging from less than 1 to 94 years old), rather than across the lifetime of each  individual. 
For a general study like the one proposed here, this could be important in smoothing out the effects on the analysis that may be due to the impact of one individual.

In the context of unsupervised clustering, we consider each row as a vector representing the expression values for a certain gene across the age, where it can potentially take a binary label with respect to the assumption on its association with ageing. However, for the semi-supervised approach, we have the label for a small number of genes known to be associated with ageing, and we seek to annotate the remaining genes.  

We also used a database reported by Tacutu et al  \cite{b14} which listed 307 genes that are known to be associated with ageing, and 243 genes that are most likely associated with aging. 

   

 \vspace{-.2cm}
\section{ Unsupervised Approach}
\vspace{-.2cm}
 Unsupervised learning as a branch of machine learning that is established based on the fact that data points are generated by natural sources with limited complexity, rather than random sources. It essentially  deals with problem domains, such as data clustering, annotation, and dimensionality reduction \cite{b24,b25}. One of the well-known unsupervised learning algorithms is $k$-means clustering, where the goal is to assign the data points to $K$ clusters in an iterative fashion \cite{b26,b34}. 
 
 In this work, 
 before applying the $k$-means algorithm, we take an information-theoretic approach, and compute a few measurements such as  entropy, correlation and K-L divergence, and represent vector of expression values corresponding to each gene with these measurements. In fact, extraction of these  measurements represent a  limited form of differential gene expression over the age range. Thus, the measurements are expected to guide the $k$-means algorithm to more directly find any potential associations between the genes and the ageing process. Although we will also represent the result of applying K-Means on the raw data to ponder upon the effectiveness of using these measurements. 
 
 Recent studies suggest that less than 2${\%}$ of genes in the human skin, retina and blood leucocytes are identified with age-related changes in expression level \cite{b1,b9,b10,b11}. Accordingly, since we have a huge number of genes and we want to select only a small subset of genes, we need to eliminate many of them through clustering. Therefore, the main idea is to eliminate most of the genes that are deemed not to  contribute to the ageing process. 
 Along this line, three measurements, namely entropy, correlation, and Kullback-Leibler divergence are employed to discriminate the genes in terms of their association with ageing. In the next few sections, we will discuss these measurements and how we employ them to represent the data, before applying $k$-means algorithm. After considering 
considering each measurement separately, we combine the individual results using ensemble modeling.

 
We expect that genes related to ageing or longevity, will typically influence aging, or be affected by aging, in a manner that is consistent over time, along the age axis. Therefore, we divide the expression values of each gene into two age groups, generally representing young subjects and old subjects respectively, and then examine the identified  measurements across  all the subjects, within each age group, and between the groups.  
 To find the appropriate age threshold to divide the subjects into binary groups, we performed  experiments using different ages, namely,  25, 30, 35, 40, and 45. Then we compare the results to determine the best age threshold.
 
\subsection{Entropy}
We compute the entropy of each gene across all 143 subjects and also across the pair of age groups for each age threshold.

The entropy of a random variable was first defined by Claude Shannon in 1948 as the average amount of information or uncertainty in the variable's possible outcomes 
\cite{b36}.
In information theory literature, it is represented as follows:
\begin{equation}\begin{split}
E = \sum{-p_i}\log{p_i}
\end{split}
\label{eq:1}
\end{equation}
where ${p_i}$ is the probability 
of the $i$-th outcome.
However, we need to compute the entropy of histogram of expression values for each gene as the entropy of the probability distribution of the values within the age bins of the histogram, in which ${p_i}$  is defined as the number of values within a given age bin divided by the total number of expression values for that gene \cite{b29}. Here we empirically set the number of histogram bins to 7, after testing different values.  
We select the genes based on their corresponding contribution to aging, as measured by the entropy of expression values for  each gene across all the subjects, or within the two separate groups of subjects for a given age threshold, for instance,   \textbf{Group 1} (1 to 25 years old) and \textbf{Group 2} (26 to 94 years old) using an age threshold of  25. In fact, posing the gene selection problem within the context of age groups allows for a dynamic consideration in the selection process. That is, if gene functionality depends on the age of the subject,  then, we expect the model to be able to  recognize it. Also, by using  multiple age thresholds, it becomes possible to see which genes play more significant roles at different stages in life, as captured by different age ranges.   

\textbf{Entropy Across All Subjects. } As the first scenario, we compute the entropy of genes individually across all 143 subjects and then select the top $T$ 
genes with the highest entropy. The intuition behind this is that the genes conveying more information through their expression values across age range of 1 to 94, are more probable to be related to the ageing process.

\textbf{Entropy Across Subject Groups. }
As the second scenario, we divide the subjects into two groups for a given age threshold, based on their age. For instance, using an age threshold of 35 we will group the subjects into two age groups: Group 1: ages [1-35], and Group 2: ages [36-94]. Then the entropy of the genes across the subjects in each age group are computed individually and top $T$ genes are selected based on their entropy ranking. Thus, each age group will produce one gene ranking.


\textbf{Entropy Difference. }
Given the binary grouping, for a given gene it becomes possible to consider its  entropy difference across the subjects in the two groups. Thus, as part of the feature measurements, we also compute  the entropy between the expression values for subjects in Group 1 and Group 2. We rank the genes based on this difference, and select the top $T$ genes for further consideration. 

For our experiments, we have set $T$=1000.

\subsection{Spearman's Correlation}
Correlation is another feasible metric for 
exploring  the underlying pattern of relationship between two variables -- in this case, between subject age and expression values of each gene from the individual. In other words, correlation can be used to measure the possible association between expression values of the genes and the ageing process. The outcome of the correlation between two variables could be positive, neutral, or negative. With positive correlation both variables change in the same direction; with neutral correlation there is no relationship in the change patterns of the variables; and with negative correlation the variables change in opposite directions. Since we are seeking for any type of age-association, we use the absolute value of the correlation for further computation and inference.

There are two formulae for computing the correlation, namely Pearson’s correlation and Spearman’s correlation. The Pearson correlation coefficient can be used to measure the strength of the linear relationship between two variables when  both variables follow a Gaussian or Gaussian-like distribution. While Spearman's correlation can be used for two variables with a nonlinear relationship between them, and the variables  may have a non-Gaussian distribution.

Since we are unsure of the distribution and possible relationships between these two variables, gene expression values and age, we need to use Spearman's correlation  as our metric to select the genes with most contribution. The formulation for Spearman's correlation coefficient is presented as follows:

\begin{equation}\begin{split}
L = \frac{C(R(x),R(y))}{S(x)S(y)},
\end{split}
\label{eq:2}
\end{equation}
where ${x}$ and ${y}$ are our variables, ${R(x)}$ and ${R(x)}$ are rank of variable ${x}$ and ${y}$, ${S}$ is the standard deviation and ${C}$ represent the notation for covarience.





Similar to entropy, for each gene we also compute the Spearman's correlation between the expression values and age, for all subjects, within each of the two age groups, and the difference in the correlation for subjects across the two age groups.

\subsection{Kullback-Leibler Divergence}
\label{sec:KL}

The Kullback-Leibler (KL) divergence (also called  relative entropy) measures  the difference between two probability distributions \cite{b36}. 
Suppose we have two discrete probability distributions ${P}$ and ${Q}$ which are defined on the same probability space, ${Z}$, the Kullback–Leibler divergence from ${Q}$ to ${P}$ is defined as follows: 
\begin{equation}\begin{split}
D_{KL}(P||Q) = \sum_{z  \in  Z}{{P(z)}\log{\frac{P(z)}{Q(z)}}},
\end{split}
\label{eq:3}
\end{equation}
For our purposes, given the age threshold, for each gene we compute the KL divergence for the two groups of expression values (as determined by the given age threshold). The genes are then ranked by their KL-divergence over the two groups, and we select the top $T$ genes with maximum divergence.

Given the information-theoretic measurements, we can combine the individual top $T$ rankings, to obtain one integrated set of genes for more refined analysis, for instance, using semi-supervised machine learning methods. For instance, we can combine the tip $T$ genes, by using simple intersection or union for the individual top-$T$ ranked gene sets. An alternative approach is to apply further unsupervised learning techniques on the data, for instance, using $k$-means clustering, but this time using the information-theoretic measurements as the features.  The clustering results will then be passed to the semi-supervised learning algorithm for final refinement of the identified genes. In this work, we show results for the two approaches.
\subsection{Union of Selected Genes Using Measurements}
As it is presented in Table I, the measurements provide different performance in gene selection, where the best performance is associated with K-L divergence. This could be due to the fact that each gene relationship with aging would only be capture by only one a few of measurements, and not all of them. Considering this, one way to select a subset of genes from 27142 genes and pass them to next stage (semi-supervised approach) is two find the union between top genes selected by each measurement as the mentioned subset. The result of this union of top 1000 genes selected by each of the nine measurements in Table I provides us with 6799 genes. We pass it to semi-supervised gene selection stage where the results are presented in Table II.

\subsection{K-means Clustering}
A popular clustering method is the $k-means$ algorithm. Though this algorithm is well known, given the significance of clustering in our approach, we briefly describe the algorithm below, for completeness.
 The $k$-means clustering algorithm, takes a training set ${x^{(1)}, ... , x^{(m)}}$, and the goal is to group the data points into a few predefined number of coherent clusters. To describe a customised version of the algorithm for this work, say we are given feature vectors for each gene (data point) $x^{(i)} \in \mathbb{R}^n$ where $n$ is the dimension of the  feature vector. The algorithm initially selects $k$ (here k=2) random data points as the centroid of the $k$ clusters, and then according to the distance of the remaining data points to each of these centroids, assigns a cluster to the data point. Next, the mean of each cluster is computed as the new centroid of the cluster, and the update process of assigning clusters to the data points repeats until all data points remain at the assigned centroids and consequently the centroids could not be updated anymore.
 
 More specifically, here rather than 143 measurements, we represent genes individually by nine information-theoretic measurements as described above,  and apply ${k}$-means algorithm on the gene expression dataset to assign each of the 27,142 genes to one of the binary  clusters. As explained eariler, we 
  here we set $k=2$ since we wish to classify the genes as being associated with age, or otherwise. Thus, the goal is to predict $k=2$ centroids and a label $c^{(i)}$ for each gene as our data points. The k-means clustering algorithm is as follows:\\

1.Randomly initialize the cluster centroids ${\mu_1}$ and ${\mu_2}$.

2. Repeat until convergence ${\{}$

For each ${i\in\{1, 2, ..., 27142\}}$, set:

\indent \indent \indent \indent \indent ${c^{(i)}=arg min_j ||x^{(i)}-\mu_j||^2}$

For every ${j\in \{1, 2\}}$, set:

\indent \indent \indent \indent \indent${\mu_{j}=\frac{\sum_{i=1}^{M}1\{c^{(i)}=j\}x^{(i)}}{\sum_{i=1}^{M} 1 \{c^{(i)}=j\}}}$ ${\}}$


\section{Semi-supervised Approach}
\vspace{-.2cm}
Unsupervised gene selection, provided us with two clusters of genes, which beside the theoretical support, we experimentally examined that the smaller cluster, with 5669 genes, includes previously known genes, and hence makes a good case for further refinement for gene selection. Thus,  we seek to further analyse this cluster to select a small subset of genes using a semi-supervised approach. 
In order to efficiently and reliably identify new genes associated with ageing, we need to leverage our prior knowledge, i.e., we use 'previously identified subset of age-related genes' in the literature. Here the central idea is to use the information of vector of expression values from each of 307 identified genes to select a new subset of age-related genes. Therefore, we need an appropriate  measurement of the similarity between the vector of expression values for each gene in the cluster with 5669 genes and the set of 307 known genes. 
Mutual information could be one good metric to find the most similar genes to the set of known genes. However, mutual information between two variables requires an exact or at least approximate estimation of joint probability distribution. However, due to the lack of further information such as multiple vectors rather than one vector of expression values for each gene, we can not compute or approximate the joint probability distribution. Instead, we use Jensen-Shannon divergence as a probability-based similarity measurement, which is based on the K–L divergence, except unlike K-L divergence, it always has a finite value and it is symmetric. On the other hand, in contrast to  the probability-based similarity measurement, we also use an efficient sample-based similarity measurement to explore the similarity both from probabilistic point of view and sample-based point of view. Cosine similarity as a sample-based similarity measurement, could be a feasible measurement of the similarity in features between two vectors of expression values in this case. 

However, one should consider the fact that these 307 known genes, obviously would not all function in the same way. For instance, some of the genes as the cause of ageing, would associate with ageing, while some of them are those genes that are just influenced by ageing \cite{b1}, and some of them might even indirectly slow down the whole process of ageing. To address this, we initially analyse the probability space representing all these 307 genes, aiming at finding some sub-spaces each associated with specific type of gene functionality. This is akin to identifying anonymous gene families. 

\subsection{Sub-Space Analysis}
The goal is to perform sub-space search and clustering of the set of 307 known genes by employing $k$-means for ${K=2,3}$, and then comparing  the hypothesized age-related genes resulting from the unsupervised learning stage against the members of the respective clusters of the known genes for further similarity analysis.  As it is formerly presented, the $k$-means algorithm provides us with $k$ clusters of data points, here clusters of genes. We set number of clusters $k=2$ and $k=3$ and refine the gene selection using the Jensen-Shannon divergence (JSD) and cosine similarity. For $k=2$ as binary clustering, the $k$-means algorithm provided us with two clusters consisting of 176 and 131 known genes, respectively. For $k=3$, we had clusters consisting of 51, 117 and 139 known genes, respectively. Latter the results corresponding to each number of clusters will be assessed.


Age-related genes are not expected to all function in the same way. Some may have a causal relationship with ageing, that is, influencing  the aging process, while the function of some others will be influenced by ageing. Even so, some genes with causal relationship with ageing could speed up ageing or slow it down. Moreover, some  studies have  suggested the possibility of age-related dynamic changes in form of monotonic over-expression, monotonic under-expression, or a mixture of these  over the life span. Accordingly, our rationale behind the sub-space search is that these types of different associations, functionalities or monotonic/non-monotonic patterns of expression should manifest in terms of "different" clusters of age-related genes. Later in this work, we will empirically examine the role of this type of  subspace analysis in terms of  clustering of the known genes \cite{b18}.
\vspace{-.1cm}
\subsection{Jensen-Shannon Divergence}
\vspace{-.1cm}
Jensen-Shannon divergence (JSD) or information radius is an information-theoretic approach to measure the similarity between two probability distributions
\cite{b36}.
This similarity measurement which has been widely employed in bioinformatics and genome comparison, is represented as follows:
\begin{equation}\begin{split}
JSD(A||B) = \frac{1}{2}D(A||M) + \frac{1}{2}D(B||M),\\
 \indent M=\frac{1}{2}(A+B)
\end{split}
\label{eq:7}
\end{equation}
where ${A_i}$ and ${B_i}$ represent the probability distribution for two given vectors of gene expression data, and D (.||.) is K-L divergence as was introduced earlier in Section \ref{sec:KL}. 

\subsection{Cosine Similarity}
Cosine similarity is a well-known approach for measuring  similarity between two vectors in inner product space. With this measure of similarity, we do not need to estimate any probability distribution, we only need to measure the cosine of the angle between the two vectors. Let $\mathbf{A}$ and $\mathbf{B}$ be  two non-zero vectors of gene expression values; the cosine similarity between $A$ and $B$
is computed  using the dot product and the magnitude as follows:
\begin{equation}\begin{split}
S(\mathbf{A},\mathbf{B}) = \frac{\mathbf{A}.\mathbf{B}}{||\mathbf{A}||||\mathbf{B}||}=\frac{\sum_{i=1}^{n}A_iB_i}{\sqrt{\sum_{i=1}^{n}A_i^2}\sqrt{\sum_{i=1}^{n}B_i^2} }
\end{split}
\label{eq:22}
\end{equation}
where ${A_i}$ and ${B_i}$ represent the elements of the corresponding vectors and ${n}$ is the number of elements of each vector.
It is noteworthy to mention that we perform the cosine similarity on both raw vectors of expression data for genes (as presented in Table II and III) as well as a vectors of nine measurements for genes individually. However, we observed that the results for the raw vectors are better. This could be due to the fact that in order to extract the phase information from vectors, we should solely consider the original raw values of the vectors rather than handcrafted features of that vectors.

\subsection{Similarity to the Set of Known Genes}
The unsupervised learning stage  will produce two initial sets of selected genes, namely, 
one with 6492 (6799-307) genes using union of top $T=1000$ ranked genes, and the other one with 5362 (5669-307) genes from  $k$-means. We will then refine these initial selections using semi-supervised methods to select those that are most to the set of known genes, using cosine similarity and JSD. 

As noted earlier, age-related genes are not necessarily monolithic -- they could still cluster into different groups, depending on the basis of the clustering. Thus, we perform the  similarity based refinement for our gene selection for three scenarios based on the clustering on the known genes, namely using one cluster, two clusters, and three clusters, respectively,

\textbf{First Scenario: One Cluster}. 
To perform gene selection for this scenario, we compute the average JSD as well as average cosine similarity of each gene with all 307 genes and then select the top $Q$ genes with highest similarity values. We do this for both JSD and cosine. Accordingly we will have:
\begin{equation}\begin{split}
S_j= \sum_{i=1}^{307}\frac{|S_{i,j}|}{307}; \indent 
JSD_j= \sum_{i=1}^{307}\frac{JSD_{i,j}}{307},
\end{split}
\label{eq:31}
\end{equation}
where ${|S_j|}$ is the absolute value of similarity between gene number $j$ (out of 53622) and the whole set of 307 genes known to be age-related. Similarly, ${ JSD_j}$ is the average of the JSD between the $j$-th gene and each of the 307 known genes. We use absolute value of cosine similarity as were are interested in any type of association, whether positive or negative similarity. To select the most similar genes in the sense of expression values, we rank the genes based on cosine similarity as well as JSD separately, and select first $Q$ genes for each method. For our experiments, we set $Q$ to 500, which is close to  the ${2\%}$ of genes in the human skin presumably with gene expression changes associated with ageing \cite{b1,b10}.

\textbf{Second Scenario: Two Clusters.} 

\vspace{-1em}
 \begin{equation}\begin{split}
S_{jc1}= \sum_{i=1}^{176}\frac{|S_{i,j}|}{176}, \indent S_{jc2}= \sum_{i=1}^{131}\frac{|S_{i,j}|}{131}; \\ 
S_j = \max{\{S_{jc1},S_{jc2}\}}; \\
JSD_{jc1}= \sum_{i=1}^{176}\frac{|JSD_{i,j}|}{176},
\indent  JSD_{jc2}= \sum_{i=1}^{131}\frac{JSD_{i,j}}{131}; \\ 
JSD_j = \max{\{JSD_{jc1},S_{jc2}\}}
\vspace{-0.5em}
\end{split}
\label{eq:3b}
\end{equation} 
\vspace{-0.15em}
 where ${S_{jc1}}$ and ${S_{jc2}}$ are the similarity between genes number $j$ (out of 5452) and cluster 1 with 176 genes and cluster  2 with 131 genes, respectively.
 Next, we rank the genes separately, based on ${S_{j}}$, and ${JSD_{j}}$, respectively. Then, for each ranking, we select the top $Q$ genes. 
 
 \textbf{Third Scenario: Three Clusters.} Similarly, we perform clustering to divide the 307 genes into 3 clusters, and compute the similarity values:
  \begin{equation}\begin{split}
\noindent S_{jc1}= \sum_{i=1}^{139}\frac{|S_{i,j}|}{139},\indent S_{jc2}= \sum_{i=1}^{117}\frac{|S_{i,j}|}{117},
\indent S_{jc3}= \sum_{i=1}^{51}\frac{|S_{i,j}|}{51}; \\ 
S_j = \max{\{S_{jc1},S_{jc2},S_{jc3}\}}. \indent
\text {Similarly, we compute:}\\
JSD_j = \max{\{JSD_{jc1},JSD_{jc2},JSD_{jc3}\}}
\end{split}
\label{eq:3c}
\end{equation}
where ${S_{jc1}}$, ${S_{jc2}}$ and ${S_{jc3}}$are the similarity between genes number $j$ (out of 5452) and clusters  1, 2 and 3, respectively. We select top $Q$ ranked genes separately, based on ${S_{j}}$ and  ${JSD_j}$ respectively.

\section{Results and Performance Evaluation}
\label{results}

In this section, we evaluate the performance of the proposed methods, for both supervised and  semi-supervised approaches. Results of the unsupervised approaches are used as the input to the semi-supervised learning approach for  final gene selection. 

\subsection{Evaluation of Unsupervised Approach}
In order to evaluate the unsupervised clustering approach, we contrast the results with the list of the 307 genes  \cite{b14} known to be 
associated with aging. The results using each of the measurements are presented in  Table I.
As the table shows, clearly, KL divergence provided the best performance -- recognizing  254 genes out of 307 known age-related genes, within its top $T=1000$ ranked genes. 
 The  best age threshold appears to be at age 40, followed by age 35. Incidentally, these are quite close to the age range 38 to 40, which Belsky et al \cite{b32} found to be significant in considering the pace of ageing in young adults. Rahman and Adjeroh \cite{b33} also observed that human physical activity and various biomarkers of biological aging show a turning point around this age range. These show that the age range 38 to 40 could represent a major dividing point for most individuals in terms of age progression. 
 
 
\begin{table}
\begin{center}
\caption{Performance of each  measurement in ranking the genes with respect to association with aging, at a given age threshold ($\tau_A$). Entries show number of the 307 known age-related genes that are  recognized within the top $T= 1000$  genes selected by each measurement at the indicated age threshold.  
}
\begin{tabular}{|p{0.1cm}|p{2.43cm}|p{.65cm}|p{.65cm}|p{.65cm}|p{.65cm}|p{.65cm}|}
\hline
${\#}$ & Measurement & $\tau_A= 25$ & $\tau_A= 30$ & $\tau_A= 35$ & $\tau_A= 40$ & $\tau_A= 45$\\

\hline\hline
 1 & Entropy (all subjects)& 232 & 232 &  232 & 232& 232\\ \hline
 2 & Entropy (Group 1) & 112& 166& 201  & \textbf{207} & 132\\
\hline
 3 & Entropy (Group 2)  & 108 & 118 &  \textbf{196} & 192& 127\\
 \hline
 4 & Entropy difference  & 128 & 124 & 138 & \textbf{162} & 125\\
\hline
  5 & Corr. (all subjects)  & 214 & 214&  214& 214 &214\\
\hline
 6 & Corr. (Group 1) & 79& 125& 177 & \textbf{183} & 155\\
 \hline
 7 & Corr. (Group 2) & 112 & 108 & 128 & \textbf{163} & 125\\
 \hline
 8 & Corr. difference & 83 & 103 & 116 & 129 & \textbf{132}\\
\hline
9 & K-L Divergence & 201 & 209& 239 & \textbf{254} & 233\\
\hline
\end{tabular}
\end{center}

\label{tab:res_ageThresholds}
\end{table}
Besides, the $k$-means algorithm provides us with two clusters of 5669 and 21,473 genes respectively. Although the second cluster contains even genes with all measurements equal to zero, the first cluster (5669 genes) contains all the genes that are known to be age-related. Therefore, we consider this smaller cluster as the one that most probably would contain other age-related genes that are  not yet known. Hence, we pass this cluster to the semi-supervised algorithm to leverage knowledge from known genes to further refine the selection of new genes that are most likely to be age-related.

\subsection{Results of Semi-supervised Learning }
To this end, besides the list of 307 known genes used to train  the semi-supervised learning algorithm, we use an additional 243 genes also known to be associated with aging to test our proposed methods for identifying novel genes with a high likelihood of association with  aging.

Table \ref{tab:res_geneRanking_asInput} shows the results using the combined top $T$ ranked genes from the individual information-theoretic measurements, using unsupervised learning as input to the semi-supervised method. With $T$ set at 1000, taking a union of all the top $T$ ranked genes (at  $\tau_A=40$, see Table \ref{tab:res_ageThresholds}) resulted in a total of 6799 unique genes, which were  then fed to the semi-supervised learning algorithm. The  combined cosine distance and JSD results were obtained by simply computing  the average rank from the two methods. 


\begin{table}
\caption{Results using the union of the individual top $T=1000$ ranked genes from the nine measurements as input to the refinement stage with semi-supervised learning. Results show the number of the 243 known genes found in the top $Q=500$ selected genes  using cosine similarity, JSD, and their combination in three scenarios of clustering on 307 known genes. 
}
\begin{center}
\begin{tabular}{|c|c|c|c|}
\hline
Metric & 1 cluster & 2 clusters & 3 clusters \\
\hline\hline
 Cosine & 200 (${82.3\%}$) & \textbf{211 (${86.8\%}$)}  & 169 (${69.54\%}$)  \\ 
 \hline
 JSD &  \textbf{221 (${90.9\%}$)} & 218 (${89.7\%}$)  & 178 (${73.2\%}$)  \\
\hline
 Combined & 218 (${89.7\%}$) & \textbf{223 (${91.7\%}$)} & 163 (${67\%}$) \\
\hline

\end{tabular}
\end{center}

\label{tab:res_geneRanking_asInput}

\end{table}
\vspace{-.2cm}

\begin{table}
\caption{Results using the cluster of selected genes as as input to the refinement stage with semi-supervised learning. Similar to Table  \ref{tab:res_geneRanking_asInput}, results show
the number of the 243 known genes found in the top $Q=500$ selected genes. 
}
\begin{center}
\begin{tabular}{|c|c|c|c|}
\hline
Metric & 1 cluster & 2 clusters & 3 clusters \\
\hline\hline
 Cosine & 228 (${93.8\%}$) & \textbf{238 (${97.9\%}$)}  & 199 (${81.8\%}$)  \\ 
 \hline
 JSD &  \textbf{231 (${95\%}$)} & 223 (${91.7\%}$)  & 165 (${67.9\%}$)  \\
\hline
 Combined & 234 (${96.2\%}$) & \textbf{243 (${100\%}$)} & 188 (${77.3\%}$) \\
\hline

\end{tabular}
\end{center}

\label{tab:res_geneClusters_asInput}
\end{table}

For the semi-supervised scenarios using different number of clusters, the results show that clustering using $k=2$ clusters tend to lead to better  results for similarity measurements.  Tables \ref{tab:res_geneRanking_asInput}
and \ref{tab:res_geneClusters_asInput}
show that final selection refinement using inputs as the gene clusters based on clustering in the high-dimensional feature space performed better that using the union of the top $T$ ranked genes from individual information theoretic measurements.  
The results indicate that, at $k=2$ clusters, the top $Q=500$ identified genes contained most, if not all of the 243 known genes. Thus, each method identified over 250 other genes that are potentially associated with aging. 
Further analysis showed that, aside from the 243 known genes, taking the intersection of the results from the two approaches (combined results using $k=2$ clusters) resulted in a total of 82 unique genes. Thus, these  represent novel age-related genes identified by our proposed approach.




\section{Discussion and Conclusion}
\label{conclusion}
\vspace{-.2cm}
As  presented in Table I, one can easily contrast the result of each of the measurements independently. As it is clear, based on the recognition of known genes among top 1000 selected genes, K-L divergence and entropy across all subjects presented the best selection performance. 

On the other hand, in order to assess the effectiveness of representing each gene with nine measurements to $k$-means rather than utilizing raw vector of gene expression values as the input to $k$-means algorithm, we perform the $k$-means on the raw vector of gene expression values in order to cluster the whole set of 27142 genes into two cluster.
The result of $k$-means clustering on the raw data set, presents two clusters with 9479 and 17633 genes respectively. The second cluster consists of genes with no expression values along the age range, while the smaller cluster apparently is more associated with ageing as it contains known genes. As this cluster (9479 genes) represents more genes than the cluster (5669 genes) attained by $k$-Means on measurements, one might consider it less reliable than then smaller cluster. However, we passed this cluster to the semi-supervised method for further investigation. The top 500 genes picked based on cosine similarity to the set of known genes, contain only ${56\%}$ of the 243 remaining known genes. This strongly implies the advantage of the $k$-means on the measurements (as our preprocessing step) over $k$-means on the raw gene expression data. 


In this work, we present and apply a computational approach using an information-theoretic framework on a dataset of human dermal fibroblast gene expression data, in order to identify novel genes associated with human ageing. First we apply information-theoretic measurements on the data to compute important features for our analysis. We then rank the genes based on these measurements, and perform binary clustering on the genes. Next, we leverage  prior knowledge about the subset of genes already known to be age-related to further refine  the initially identified  genes in the clusters. We presented results showing the performance assessment of the proposed framework, indicating  the accuracy and effectiveness of  the approach. The results could be further improved with availability of more known age-related genes. Yet, our results could be used to further guide  laboratory experiment studies on the mechanisms of ageing and age-related genes. 

\vspace{12pt}

\end{document}